# OBSERVATIONS OF GAS AND DUST IN THE GALACTIC CENTRE

**Glenn J. White [1], H.A. Smith [2], G. Stacey [3], C. Matt Bradford [4], S.J. Leeks [5], O. Sweet [1], J. Fischer [6]**

(1) School for Physical Sciences, University of Kent, Canterbury CT2 7NR, England g.j.white@kent.ac.uk (2) Harvard-Smithsonian Center for Astrophysics (3) Department of Astronomy, Cornell University, (4) \Department of Physics, Caltech , (5) Department of Physics, University of Cardiff, (6) Naval Research Lab, Washington D.C. USA

**Abstract**

The Infrared Space Observatory (ISO) Long wavelength Spectrometer (LWS) has been used to map distribution of the emission from a sample of 22 atomic, molecular and ionised lines toward the Circumnuclear Disk (CND) at the Galactic Centre. The CND disc is clearly seen in the maps of molecular lines such as CO and OH, whilst the central region dominates in other atomic and ionised lines such as OIII and NIII. The spectrum toward Sgr A* is best represented by the sum of a 58 K blackbody, superposed with 22 identifiable emission or absorption features, including four lines each attributed to CO and OH, two broad features that may be indicative of a complex of solid state features, two $H_2O$ lines, and the rest being various atomic or ionised atomic lines. The CO emission is best fit by gas having $T_{kin}$ ~ 900 - 1400 K and $n_{H2}$ ~ $10^4 - 10^{4.5}$ cm$^{-3}$.

**Introduction**

The nucleus of our Galaxy is located ~ 8.5 kpc from the Earth, more one hundredth of the distance to the nearest large external galaxy, or one thousandth that to the nearest active galaxy. Due to relatively high spatial resolution achievable, it provides a unique insight into the structures of more distant nuclei and is our nearest template for the conditions that pust exist in many other similar galactic nuclei. The distribution of atomic molecular and ionised gas toward the Galactic Centre (GC) has been summarised by many authors, including Genzel Townes (1987), Genzel, Hollenbach Townes (1994), Shields & Ferland (1994), Zylka et al (1994, 1995), and Coil & Ho (2000). An unusual compact (size ~ 1 AU - Krichbaum et al 1993, Rogers et al 1994) variable radio source lying close to the dynamical centre of the Galaxy, Sgr A* has been suggested to be either a quiescent black hole. Within 1″ (0.2 pc) of Sgr A* is the bright infrared complex, IRS16, which contains at least 15 components, and is probably the source of the strong stellar wind which has velocities of up to ≥ 700 km s$^{-1}$. Surrounding this is a cavity - dominated by atomic and ionised gas, containing a mass ≥ 2 10$^6$ M$_\odot$, and luminosity ~ 7 10$^6$ - 2.3 10$^7$ L$_\odot$. Most of this energy is absorbed by the surrounding material, and then re-emitted as IR continuum, and by atomic and molecular line emission. Zylka et al (1995) show the column density of warm (T ~ 200 - 400 K) absorbing dust in the centre of the cavity is less than that of the surrounding CND molecular gas and dust torus.

The CND is a ring of gas and dust, rotating around the central ionised cavity with a velocity ~ 110 km s$^{-1}$ and having a mass ~ $10^4$ M$_\odot$ of clumpy material. First recognised from the far-infrared observations by Becklin, Gatley & Werner (1982) it extends from a radius of 1.5 - 2 pc out to ~ 6 - 8 pc, and may be connected to the gravitational potential of material in the GC (Duschl 1989). Molecular line studies shows the CND to be composed predominantly of highly excited gas, co-rotating in the same general direction as the rest of the Galaxy (positive $v_{lsr}$ at positive $l$, inclined at 20 – 30$^o$ to the Galactic Plane, and 20 – 25$^o$ to the line of sight - Fukui et al 1977, Genzel et al 1982, Harris et al 1985, Sandqvist & Loren 1985, Serabyn et al 1986, Güsten et al 1987 a,b). Although densest at the inner edge (radial distances ~ 2 pc), it extends at least ~ 8 pc north and south of Sgr A*. It may be a transient feature, where material in the Sgr A East dust shell has interacted with the central stellar cluster. The neutral material in this ring is believed to be hot (T ~ 150-450K), dense ($n_{H2}$ ~ $10^{3-7}$ cm$^{-3}$), and fragmented (Genzel et al 1982, 1988 - with a filling factor ~ 10%, Zylka et al 1995, McGary et al 2001, Wright et al 2001). The ring contains ≥ $10^4$ M$_\odot$ of gas (Mezger et al 1989), based on 1.3 mm continuum and CO observations. It has a clumpy and filamentary (Yusef-Zadeh et al 2001) structure that allows ultraviolet radiation from the central ionised region to penetrate into the neutral ring, heating and photoionising the gas. The likely source of this radiation is the cluster of hot, luminous stars, some of which are the HeI/HI stars detected by Krabbe et al (1991). The presence of hot gas was independently inferred from the detection of near-IR $H_2$ emission, which suggested that the material was shock excited at the edge of the ionised region (Gatley et al 1984). The neutral gas in the ring appears to be dynamically coupled to the gas in the central cavity of ionised and atomic gas, and has its rotational axis close to that of the larger scale Galactic rotation. The sharp edge of the CND abuts directly against the ionised arc-like filaments (The Mini-spiral), which has led to speculation that material from the CND may be falling towards the Centre along the ionised arms of the mini-spiral.

Mid-IR and submillimetre continuum data (Zylka et al 1995, Latvakoski et al 1999) show that the central ~ 30″ diameter cavity contains ~ 400 M$_\odot$ of dust at T ~ 40 K, 4 M$_\odot$ at ~ 170 K and ~ 0.01 M$_\odot$ at ~ 400K. The Galactic Centre is viewed through a foreground visual extinction, A$_v$ ~ 31 magnitudes (Scoville et al 2003). The luminosity of the emitted radiation for these three phases is ~ $2 \times 10^6$ $L_\odot$, $4 \times 10^6$ $L_\odot$ and $5 \times 10^5$ $L_\odot$ respectively. The hottest gas lies close to Sgr A*, whilst the brightest far-infrared emission comes from the transition region between the neutral CND and surrounding HII regions such as Sgr A East (a synchrotron shell-like source lying behind the central HII region (Pedlar et al 1989) - probably formed following an explosive event inside its core, and Sgr A West - the central HII region within which Sgr A* is embedded (Dent et al 1993). Lying further out are two molecular clouds, M-0.02-0.07 (also known as the 50 km s$^{-1}$ cloud) and M-0.13-0.08 (also known as the 20 km s$^{-1}$ cloud). These two clouds are believed to be connected together with the core of Sgr A East (Mezger et al 1989, Okumura et al 1991, Ho 1994).

**Observations**

In this paper we report new observations of the CND region around Sgr A*, including a 40 position, half beam sampled map made using the Long wavelength spectrometer (LWS - spectral range 45-195 µm) on the Infrared Space Observatory (ISO) satellite. This is supported by observations of submillimetre wavelength continuum and line emission obtained with the James Clerk Maxwell Telescope (JCMT) in Hawaii. The grid observed with ISO covered the complete CNR region, and the nearby Sgr A West source. The observations were made in the LWS full range grating scan observing mode, and were analysed using the ISAP 1.6 / LIA 1.7 standard software tools. In addition to the normal data product processing, we applied an additional strong source correction, designed to correct for non-linear behaviour in the detector response. Fig. 1 shows the ISO LWS spectrum observed towards Sgr A*. The spectrum is dominated by the 63 mm OI line, with a further 21 spectral line features that can be distinguished - some of which are blended with other fainter lines. Several of the lines are seen in absorption against the continuum emission - notably transitions of OH, H$_2$O and CH.

The lines are superimposed on top of the strong continuum from Sgr A*, making the weaker ones difficult to see on this scaling. A blackbody fit to the continuum emission gives a best fit dust temperature of 58 K - which is somewhat lower than the mid-infrared colour temperatures that range from 140 - 300 K estimated by Telesco et al (1996), Chan et al (1997), and Latvakoski et al (1999) with smaller beams. The 75 µm continuum emission peaks centrally at the location of Sgr A*, and shows some evidence for extended emission running SW-NE along the Galactic Plane.

The spectrum shown in Fig, 2 has had a 58 K blackbody spectrum subtracted from it to emphasise the line emission.

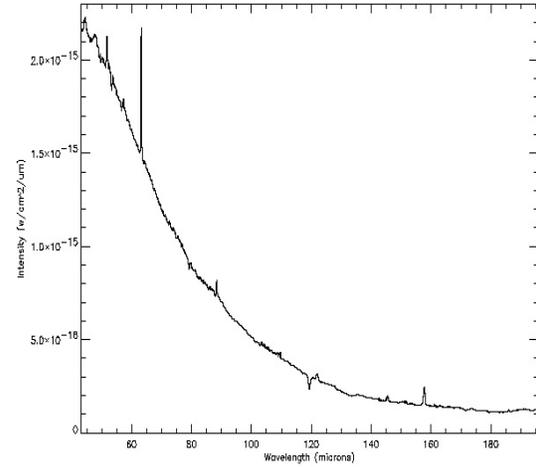

**Figure 1**. *LWS spectrum from 40 to 195 microns. The vertical scale is in units of W cm$^{-2}$ µm$^{-1}$.*

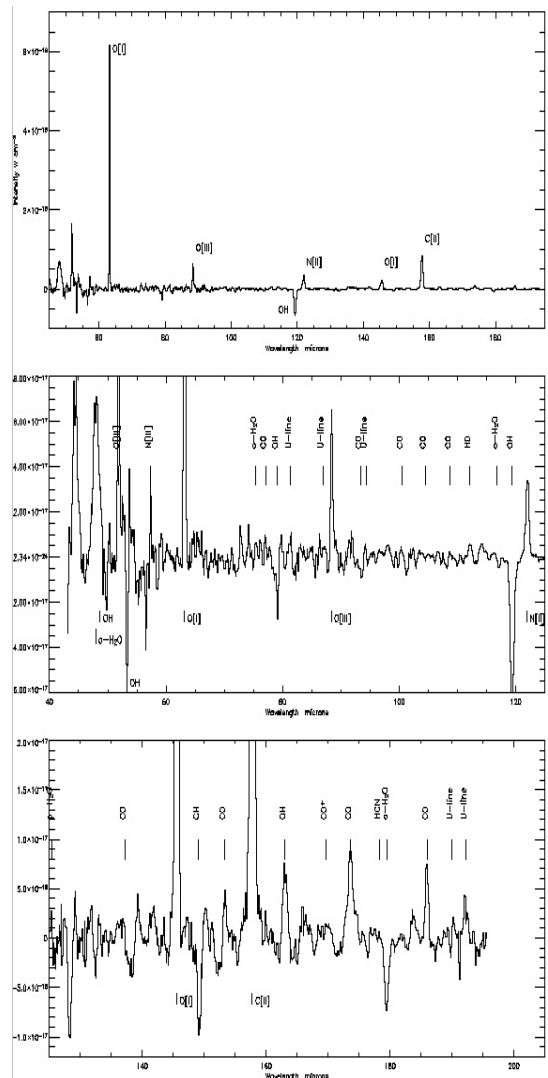

**Figure 2**. *LWS spectrum from 40 to 195 microns. With a 58K blackbody curve removed. The vertical scale is in units of W cm$^{-2}$ µm$^{-1}$.*

The CND can be clearly seen in a number of the molecular lines, including CO and OH - examples are shown in Fig 3.

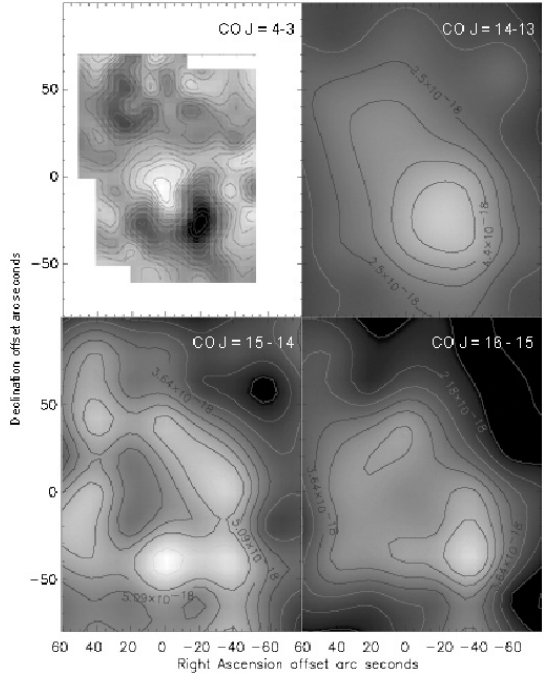

**Figure 3** *Maps in the CO J= 4-3, 14-13, 15-14 and 16-15 transition, showing the shape of the CND. The respolution of the J= 4-3 data is 10", and for the others is 70".*

We have modeled the CO transitions mapped by ISO, along with CO $J = 3-2$ and 4-3 maps obtained using the JCMT, in Fig 4. Since the CO $J = 3-2$, and $J = 4-3$ lines were observed with beam sizes of 15 and 10 arcsec respectively, it was necessary to convolve the maps with a gaussian smoothing beam to achieve the same angular resolution as the ISO data. These data were modeled using a large velocity gradient transfer code. In this model, the opacity effects in the lines are taken into account by introducing an escape probability formalism. A weak diffuse radiation field was introduced by including the thermal continuum of a dust at a temperature $T_{dust}$ = 58 K, since this was typical of blackbody fits to the shape of the continuum emission observed with the LWS. The molecular abundance of CO relative to $H_2$, X(CO), was assumed to be $8\ 10^{-5}$. The LVG code was run for a 40 level model, varying the kinetic temperature $T_{kin}$ in 100K steps between 300 and 2200K, for a range of densities $n_{H2} \sim$ between $10^3$ and $10^8$ cm$^{-3}$. The collisional rate coefficients used are described by Larsson, Liseau & Men'shchikov (2002).

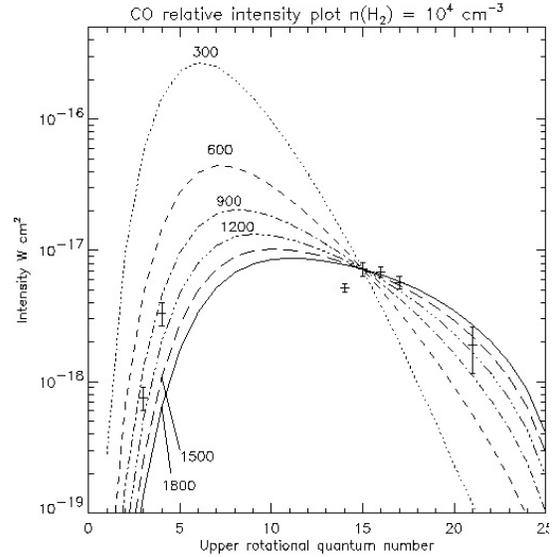

**Figure 4** *CO line intensities modeled using an LVG code, for different excitation temperatures.*

**Other lines**

Maps of the atomic and ionic emission lines are shown in Fig 5.

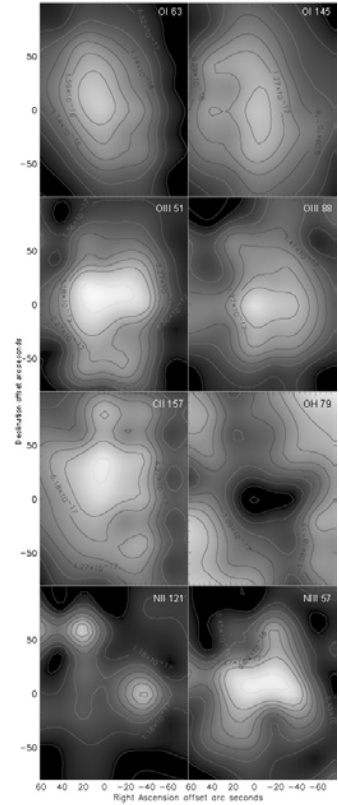

**Figure 4** *Maps of the CND running clockwise from top right; OI 145, OIII 88, OH 79, NIII 57, NII 121, CII 157, OII 51 and OI 63 µm lines.*

**Conclusions**

The spectrum toward Sgr A* is best represented by the sum of a 58 K blackbody, superposed by 22 identifiable emission or absorption features, including four lines each attributed to CO and OH, two broad features that may be indicative of a complex of solid state features, two $H_2O$ lines, and the rest being atomic or ionised atomic lines.

Maps of the CO emission show the structure iof the circumnuclear ring. The CO gas excitation is best fit by gas having $T_{kin}$ ~ 900 - 1400K and $n(H_2)$ ~ $10^4$ – $10^{4.5}$ $cm^{-3}$.

The well known molecular ring that starts ~ 2 pc from the centre is seen in lines of CO, CH and OH surrounds a central cavity which is prominent in atomic line emission from species such as carbon, oxygen and nitrogen. The CND is predominatly traced by molecular line emission, whereas the central cavity is more prominent in the atomic lines.

**References**

Aitken, D.K., Griffiths, J., Jones, B. Penman, J.M. 1976, MN, 174, 41p
Becklin, E.E., Gatley, I. and Werner, M.W. 1982, ApJ, 258, 135
Bowey, J.E., Lee, C., Tucker, C., Hofmeister, A.M., et al. 2001, MNRAS, 325, 886
Ceccarelli, C., Caux, E., White, G.J. et al 1998, A\&A 331, 372
Cernicharo, J., Lim, T., Cox, P., et al. 1997, A & A, 323, L25
Chan, Kin-Wing, Moseley, S. H., Casey, S., et al. 1997, ApJ., 283, 798
Clegg, P.E. et al. 1996, A & A, 315, L38
Coil, A.L., Ho, P.T.P. 2000, ApJ, 533, 245
Dent, W.R.F., Matthews, H.E., Wade, R. and Duncan, W.D., 1993, ApJ, 410, 650
Duschl, W.J. 1989, MNRAS, 240, 219.
Figer, D. et al, 1998, in press
Fukui, Y., Iguchi, T., Kaifu, N., et al. 1977, PASJ, 29, 643
Genzel, R., Watson, D., Townes, C., et al. 1982, in The Galactic center; Proceedings of the Workshop, Pasadena, CA, January 7, 8, 1982 (A83-40676 19-90). New York, American Institute of Physics, p. 72-76
Genzel, R., Hollenbach, D. and Townes, C.H. 1994, Rep. Prog. In Phys, in press
Genzel, R. and Townes, C.H. 1987, ARAA, 25, 377
Güsten, R., Genzel, R., Wright, M.C.H., et al. 1987, ApJ, 318, 124
Harris, A.I., Jaffe, D.T., Silber, M. and Genzel, R. 1985, ApJ Lett, 294, L93
P.T.P. Ho, in Astronomy with Millimeter and Submillimeter Wave Interferometry, IAU Colloquium 140, Astronomical Society of the Pacific Conference Series, vol. 59, eds. M. Ishiguro and W. J. Welch, p. 161. San Francisco: Astronomical Society of the Pacific, 1994.
Holland, W. S., et al. 1999, MNRAS, 303, 659
Krabbe, A., Genzel, R., Drapatz, S. and Rotaciuc, V., 1991, ApJ, 382, L19
Krichbaum, E., et al 1993, A & A, 274, L37
Lacey, J.H., Townes, C.H., Geballe, T.R., Hollenbach, D.J. 1980, ApJ, 241, 132
Latvakoski, H.M., Stacey, G.J., Gull, G.E., Hayward, T.L. 1999, ApJ, 511, 761, 1999
Liseau, R,. Ceccarelli, C., Larsson,B. et al. 1996, A\&A 315, L181
Lutz, D., Feuchtgruber, H., Genzel, et al. 1996, A & A, 315, L269
McGary, R., Coil, A. and Ho, P.T.P. 2001, ApJ, 559, 326
Mezger, P.G., Zylka, R., Salter, C.J., et al. 1989, A & A, 209, 337
Molster, F.J., Waters, L.B.F.M., Tielens, A.G.G.M. and Barlow, M.J. 2002, A & A, 382, 184
Moneti, A., Blommaert, A., Figer, D.F., Najarro, F. and Stolovy, S., 1998, this conference
Oka, T., White, Glenn J., Hasegawa, T., et al. 1999, ApJ, 515, 249
Okumura, S. K., Ishiguro, M., Kasuga, T., et al. 1991, ApJ, 378, 127O
Sandqvist, A., Wootten, A., Loren, R. B., 1985, A & A, 152 L25
Pedlar, A., Anantharamaiah, K.R., Ekers, R.D., et al. 1989, ApJ, 342, 796.
Pierce-Price, D., Richer, J.S., Greaves, J.S. et al. 2000, ApJ, 545, L121
Rogers, A.E., et al 1995, ApJL, (in press)
Scoville, N. Z., Stolovy, S.R., Rieke, M. et al. 2003, ApJ, 594, 294
Serabyn, E. and Güsten, R., Walmsley, C.M., Wink, J.E., Zylka, R., 1986, A & A, 169, 85
Shields, J.C. Ferland, G.J. 1994, ApJ, 430, 236
Swinyard, B. et al. 1996, A&A, 315, L43
White, G.J. and Padman, R.P. 1991, Nature, 354, 511
White, G. J., Astronomical Society of the Pacific Conference Series Volume 102, p171 - 178, 1976
White, G. J., Padman, R.P., Gatley, I. 2004, in preparation.
Wright, M.C., Coil, A.L., McGary, R., Ho, P.T.P. and Harris, A.I. 2001, ApJ., 551, 254
Yusef-Zadeh, F., Stolovy, S., Burton, M., Wardle, M. and Ashley, M.C.B. 2001, ApJ, 560, 749
Zylka, R., Mezger, P.G., Ward-Thompson, D., Duschl, W.J. and Lesch, H. 1995, preprint
Zylka , R., Mezger, P., Wilson, T.L. and Mauersberger, R. 1994, p 161, in The Nuclei of Normal Galaxies, edited by R. Genzel and A. I. Harris. Kluwer Academic Publishers